\documentclass[aps,amssymb,pra,showpacs,twocolumn,floatfix]{revtex4}
\usepackage{epsf}
\usepackage{amsmath}
\usepackage{graphicx}

\begin{document}

\title {Dephasing in an atom}

\author{B. Ivlev}

\affiliation
{Instituto de F\'{\i}sica, Universidad Aut\'onoma de San Luis Potos\'{\i},
San Luis Potos\'{\i}, San Luis Potos\'{\i} 78000 Mexico}


\begin{abstract}

When an atom in vacuum is near a surface of a dielectric the energy of a fluctuating electromagnetic field depends on a distance between them resulting, as known, in the force called van der Waals one. Besides this fluctuation phenomenon there is one associated with formation of a mean electric field which is equivalent to an order parameter. In this case atomic electrons are localized within atomic distances close to the atom and the total ground state energy is larger, compared to the bare atom, due to a polarization of the dielectric and a creation of a mean electric field locally distributed in the dielectric. The phenomenon strongly differs from the usual ferroelectricity and has a pure quantum origin connected with a violation of the interference due to dephasing of fluctuating electron states in the atom.

\end{abstract} \vskip 1.0cm


\maketitle
\section{Introduction}
\label{intr}
The process of quantum tunneling between classically allowed regions through a separating potential barrier is essentially modified when a tunneling particle is connected to an environment which produces an underbarrier friction \cite{LEGGETT}. The particle gives up a part of its energy to the environment and comes from under the barrier with a lower energy. The famous formalism of classical trajectories in imaginary time is used to describe the phenomenon \cite{OVCHIN1,SCHMID0,MELN,CHAKR,HANGGI1,LEGGETT1,KORSHUN,IVLEV,HANGGI2,KAGAN,OVCHIN2,WEISS}.

\begin{figure}
\includegraphics[width=6cm]{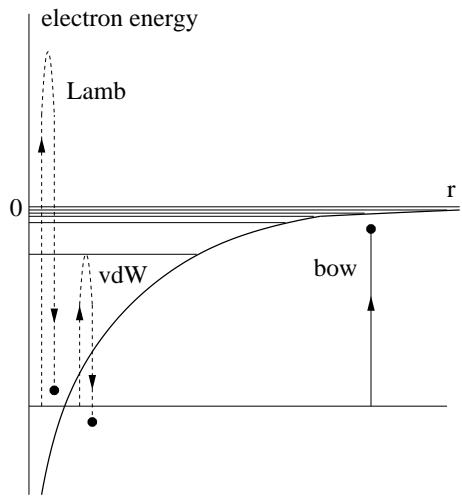}
\caption{\label{fig0}Lamb phenomenon is connected with high energies (of the order of $mc^{2}$) of virtual intermediate states. Energies of virtual states in the van der Waals phenomenon are of the atomic scale. The virtual processes are shown by the dashed curves. The electron wave function at large distances becomes a superposition containing overbarrier states which are not completely compensated by the mutual interference (dephasing).}
\end{figure}

If one classical region is a potential well and the other one is moved to the infinity one can consider eigenstates in the well of the system consisted of a particle and an environment. A typical example of such kind is an atomic electron interacting with the electromagnetic environment. Under this interaction the electron "vibrates" in the Coulomb field resulting in an elevation of the atomic energy level called the Lamb shift \cite{LANDAU3}. This process is schematically shown in Fig.~\ref{fig0} where the virtual intermediate state of the electron can
acquire the energy of $mc^{2}$.

When the atom in vacuum is close to a surface of a dielectric, electron "vibrations" produce fluctuating electromagnetic field which interacts with the dielectric. An energy of the fluctuating field depends on a distance between the atom and the dielectric resulting in a force between them called van der Waals force \cite{LANDAU1}. In this case a mean value of the electric field is also zero as in the Lamb phenomenon. This process (with respect to one atom) is shown in Fig.~\ref{fig0}.

Besides those fluctuation phenomena, a usual formation of a mean electric field in ferroelectrics is possible due to a spontaneous polarization \cite{LANDAU2}.

We argue in the paper that in the system of a dielectric and an atom close (approximately 1000$\AA$) to its surface a mean electric field can be created which plays a role of some order parameter. The underlying mechanism is completely different than the usual ferroelectricity. The electric field is formed locally in the dielectric not far from the atom. We study a ground state of the system "atomic electron + electromagnetic environment (dielectric)". The Lamb shift of atomic energy levels is also a result of an electromagnetic interaction but of a different type. In the both cases the ground state energy is higher compared to the case of a "switched off" interaction with an electromagnetic environment.

The formation of the mean field is a pure quantum effect. A dipole-dipole interaction of the atom and dielectric atoms can be considered within a perturbation theory. In each perturbation term one should sum with respect to virtual intermediate states, in particular, overbarrier waves. Every overbarrier wave is not small but the summation of oscillating functions results in their
mutual cancellation and the electron wave function remains exponentially decayed at large distances.

But since the dielectric permittivity has an imaginary part an electron motion becomes dissipative. The inelastic dephasing effects violate interference, as in localization phenomena in disordered solids \cite{KHM,GANT}. Dephasing in disordered solids and in our case equally results in reduction of the localization. The difference is that in solids {\it real} electron states interfere but in the atom {\it virtual} states participate in the interference.

Dephasing effects in the atom, violating the compensating interference, result in a less decaying electron wave function at large distances. This means a localization reduction in our case of the atom. At those distances the electron wave function becomes a superposition containing overbarrier states which are not completely compensated. In the classical language this is equivalent to an acceleration of the electron by the certain mean electric field. This can be seen if to track the underbarrier wave function along a classical trajectory in imaginary time when one can ascribe a certain energy to the electron at each point of the trajectory. In this process the electron gets more energy at large distances as shown in Fig.~\ref{fig0}.

So the reason for formation of the phase with the mean electric field is violation of the underbarrier interference (dephasing).

Accounting for dissipation of the atomic electron cannot be done within the perturbation theory and the above arguments are rather heuristic. In this situation the adequate method is the semiclassical approach used in the paper. The main steps are the following. The moving atomic electron produces an electric field which results in a dissipation since the dielectric permittivity has an imaginary part. This is similar to an electron connected to an elastic string (at some point of it) since the string can carry an energy away providing an effective friction of the electron. A free electron (disconnected from the string) moves below the Coulomb potential with the certain imaginary velocity. The finite velocity of the connected electron should be provided by the certain driving force from the string created by its elastic tension. This tension results in an additional energy of the system.

\begin{figure}
\includegraphics[width=6cm]{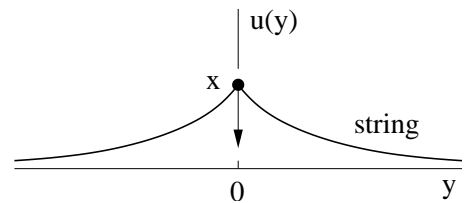}
\caption{\label{fig1}The particle of the mass $m$ is attached to the elastic string. The particle coordinate is $x=u(0)$. The string configuration results in a restoring effect on the particle.}
\end{figure}

Usually a reservoir provides a restoring effect on a tunneling particle which leads to a reduction of its energy in the underbarrier motion \cite{OVCHIN1,SCHMID0,MELN,CHAKR,HANGGI1,LEGGETT1,KORSHUN,IVLEV,HANGGI2,KAGAN,OVCHIN2,WEISS}. This situation is shown in Fig.~\ref{fig1}. In our case the string acts on the particle as a bow string on an arrow providing an anti-restoring effect of the type shown in Fig.~\ref{fig2}. This scenario can only be realized for a nonsemiclassical well with pronounced discrete levels. The bow (anti-restoring) effect does not occur in the usual case when the well is semiclassical one.

In terms of the atomic electron and the dielectric, the extra elastic energy corresponds to an energy of the locally polarized dielectric. The both phenomena, atom-dielectric and  particle-string, can be described in terms of classical trajectories in imaginary time which are of the same type for the both cases.

Creation of the mean electric field in some region of the dielectric formally reminds the analogous effect in a ferroelectric despite the dielectric itself does not exhibit any ferroelectric properties. But, as one can see, origins of those phenomena are quite different since our phenomenon is due to a delicate quantum interference. The quantum phenomenon of creation of the mean electric field also differs from a polaron formation in a crystal when the lattice is classically polarized by an electron \cite{KIT}.

When a group of atoms gets more condensed, organizing a cluster, the bow state disappears and the energy reduces. One can say that the atoms become free at small distances. This situation reminds quark confinement in particle physics since they are free at short distances (asymptotic freedom) \cite{POLYAKOV,LEE}.

In Sec.~\ref{bow} a ground state of a particle, attached to a harmonic string, is analyzed. In Sec.~\ref{atom} the theory is applied to the atom above the dielectric. In Sec.~\ref{interp} the
interpretation of the bow state is given. In Sec.~\ref{lamb} the anomalous Lamb shift of atomic levels, due to interaction with the dielectric, is considered.
\section{BOW STATE}
\label{bow}
In this section we consider a model (a particle attached to a string) when formation of the bow state occurs.
\subsection{Particle on a string}
Suppose a particle of the mass $m$ and the coordinate $x$ to be connected to the elastic string as shown in Fig.~\ref{fig1}. The string plays a role of the reservoir. The potential energy of this system is
\begin{equation}
\label{2}
U\left\{u(y)\right\}=V(x)-\hbar\sqrt{\frac{2V}{m}}\,\delta(x)+\frac{\rho s^{2}}{2}\int dy\left(\frac{\partial u}{\partial y}\right)^{2},
\end{equation}
where $x=u(0)$, $\rho$ is a mass density of the string, $s$ is a sound velocity, and $V(x)=V$ is a constant. Transverse deformations only are possible and the potential $V(x)$ is probed solely by the particle. Performing the Fourier transformation $u(y)\rightarrow u_{k}$ one can see that $u_{k}$ serve as coordinates of an infinite set of oscillators coupled to the particle as in Ref.~\cite{LEGGETT}. One can expect a renormalization of the ground state energy due to an influence of the string.

The problem is to calculate an energy of the ground state in the multi-dimensional potential (\ref{2}) where the coordinates are $u_{i}=u(y_{i})$, $i=0,\pm 1,\pm 2,..$ so that $y_{0}=0$ ($u_{0}=x$). We divide the $y$ axis by small segments. The wave function of the total system $\psi\left\{u_{i}\right\}$ corresponds to an underbarrier regime with the boundary condition $\partial\psi/\partial x=-\psi\sqrt{2mV}/\hbar$ imposed at $x=0$. An analytical solution for $\psi\left\{u_{i}\right\}$ in the whole multi-dimensional space is impossible. Nevertheless one can track the wave function $\psi\left\{u_{i}\right\}$ under the barrier along the certain trajectory $u_{i}(\tau)$, parametrized by the certain parameter $\tau$. The parameter $\tau$ has a meaning of imaginary time ($t=-i\tau$) related to a classical underbarrier trajectory which is a solution of Newton's equation in imaginary time. According to the known underbarrier scenario (see, for example, Refs.~\cite{COL,SCHMID}) the wave function decays along the trajectory and reaches a maximal value at each point of it on the surface perpendicular to the trajectory. Each particular value of $\tau$ specifies in the multi-dimensional space $\{u_{i}\}$ a point which belongs to the classical trajectory.
\subsection{Classical trajectory in imaginary time}
The classical trajectory $u(y,\tau)$ provides in the multi-dimensional space $\{u_{i}\}$ a path where the wave function is localized and exponentially decays along the path. This is a famous method to describe tunneling when a trajectory connects an initial well and a final one where a particle goes out from under a barrier. In our case there is no a final well and the particle moves from the initial well infinitely under the barrier.
\begin{figure}
\includegraphics[width=6cm]{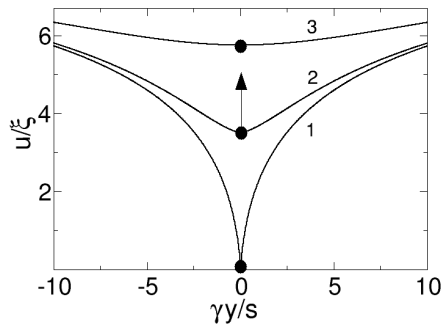}
\caption{\label{fig2}The underbarrier trajectory $u(y,\tau)$ reminds a bow string (anti-restoring effect). The solid dot is a particle position. On the curve 1, $\gamma\tau=0$, the particle energy is $E_{p}=0$, the string is most tensed with the elastic (total) energy $V$. On the curve 2, $\gamma\tau=3$, the string gives up a part of its energy to the particle. On the curve 3, $\gamma\tau=10$, the elastic and kinetic energy of the string are almost zero and the particle energy is $E_{p}\simeq V$.}
\end{figure}

Along the classical trajectory in imaginary time the total energy $E=T+U$ conserves. The kinetic energy is
\begin{equation}
\label{3}
T\left\{u(y)\right\}=-\frac{m}{2}\left(\frac{\partial x}{\partial\tau}\right)^{2}
-\frac{\rho}{2}\int dy\left(\frac{\partial u}{\partial\tau}\right)^{2}.
\end{equation}
One can obtain $E$ if to insert into Eqs.~(\ref{2}) and (\ref{3}) the classical trajectory $u(y,\tau)$ determined by the equations
\begin{equation}
\label{4}
m\frac{\partial^{2}x(\tau)}{\partial\tau^{2}}+2\rho s^{2}\frac{\partial u}{\partial y}\bigg|_{y=+0}=0,\hspace{0.3cm}
\frac{\partial^{2}u}{\partial\tau^{2}}+s^{2}\frac{\partial^{2}u}{\partial y^{2}}=0\,.
\end{equation}
In the second Eq.~(\ref{4}) $y\neq 0$. The particle coordinate is $x(\tau)=u(0,\tau)$, $\tau$ varies between 0 and infinity, and $u(y,\tau)=u(-y,\tau)$. A solution of Eqs.~(\ref{4}) has to
satisfy the conditions
\begin{equation}
\label{4a}
x(0)=0,\hspace{0.5cm}\frac{\partial x}{\partial\tau}\bigg|_{\tau=0}=\sqrt{\frac{2V}{m}}
\end{equation}
in order to account for the $\delta$ function in the potential (\ref{2}). The solution has the form
\begin{equation}
\label{6}
u(y,\tau)=\xi\int^{\infty}_{0}\frac{2\gamma d\omega}{\omega(\omega +\gamma)}\left[1-\exp\left(-\omega|y|/s\right)\cos\omega\tau\right],
\end{equation}
where 
\begin{equation}
\label{6a}
\xi=\frac{1}{\pi\gamma}\sqrt{\frac{2V}{m}}
\end{equation}
is a coherence length. The damping coefficient $\gamma=2\rho s/m$ is introduced which corresponds to the classical equation of motion
\begin{equation}
\label{5}
m\frac{\partial^{2}x}{\partial t^2}+m\gamma\frac{\partial x}{\partial t}=-V'(x)
\end{equation}
for a particle attached to a string \cite{LEGGETT}. We consider the limit of a small friction $\hbar\gamma\ll V$. The solution (\ref{6}) is plotted in Fig.~\ref{fig2} for three different values of $\tau$. An expression for $x(\tau)=u(0,\tau)$ follows from Eq.~(\ref{6}). In the limiting cases
\begin{equation}
\label{7}
x(\tau)=\xi
\begin{cases}
\pi\gamma\tau,&\tau\gamma\ll 1\\
2\ln\left(\gamma\tau\right),& 1\ll\tau\gamma.
\end{cases}
\end{equation}
In the solution (\ref{6}) $u(y,\tau)$ is positive. It tracks the wave function along the trajectory at positive $x$. Another solution of Eqs.~(\ref{4}) is negative which is just a mirror reflection with respect to the $y$ axis in Fig.~\ref{fig2}. It tracks the wave function at negative $x$.

Since $u(0,\tau)=x(\tau)$ it follows from Eq.~(\ref{6}) that
\begin{equation}
\label{7a}
\frac{\partial u(y,\tau)}{\partial y}=\frac{{\rm sgn}\,y}{\pi s}\int^{\infty}_{-\infty}\frac{\partial x(\tau_{1})}{\partial\tau_{1}}\,\frac{(\tau_{1}-\tau)d\tau_{1}}{(\tau_{1}-\tau)^{2}+y^{2}/s^{2}}.
\end{equation}
By means of Eqs.~(\ref{4}) and (\ref{7a}) one can obtain the equation for $x(\tau)$
\begin{equation}
\label{7b}
-m\frac{\partial^{2}x(\tau)}{\partial\tau^{2}}
+\frac{m\gamma}{\pi}\int^{\infty}_{-\infty}\frac{\partial x(\tau_{1})}{\partial\tau_{1}}\,\frac{d\tau_{1}}{\tau-\tau_{1}}=-V'(x),
\end{equation}
where the integral has its principle value and we put a general $V(x)$ in Eq.~(\ref{2}). A solution should satisfy the conditions (\ref{4a}). In our case the potential force is zero. Eq.~(\ref{7b}) coinsides with one of Caldeira and Leggett \cite{LEGGETT}. Instead of a solution of Eqs.~(\ref{4}) one can solve Eq.~(\ref{7b}) with the conditions (\ref{4a}) and then to find the entire $u(y,\tau)$ using a relation of the type (\ref{7a}).
\subsection{Underbarrier wave function}
Each curve in Fig.~\ref{fig2} represents in the multi-dimensional space $\{u_{i}\}$ a point which belongs to the classical trajectory. Under sweeping of $\tau$ we move along the trajectory or, in other words, along the valley of the maximal wave function. At each point of the trajectory $|\psi(\tau)|^{2}\sim\exp\left[-A(\tau)\right]$ where the classical action has the form
\cite{LANDAU,COL}
\begin{equation}
\label{9}
A(\tau)=\frac{2}{\hbar}\int^{\tau}_{0}d\tau_{1}\left(-T+U-E\right)=-\frac{4}{\hbar}\int^{\tau}_{0}d\tau_{1}T(\tau_{1}).
\end{equation}

One can easily evaluate $A(\tau)$ by inserting the solution (\ref{6}) into Eq.~(\ref{9}). We should know the wave function on the trajectory not as a function of $\tau$ but as a function of physical coordinate $x$. For this reason, $\tau$ should be expressed through $x$ from the relations (\ref{6}) and (\ref{7}). As follows from Eq.~(\ref{6})
\begin{equation}
\label{10}
|\psi(x)|^{2}\sim
\begin{cases}
\exp\left[-A_{WKB}(x)\right];& |x|\ll\xi\\
\exp\left[-A_{WKB}(x)/2\right];& \xi\ll |x|,
\end{cases}
\end{equation}
where the conventional WKB action $A_{WKB}(x)=\left(2|x|/\hbar\right)\sqrt{2mV}$ corresponds to a nondissipative ($\rho=0$) case \cite{LANDAU}. At a short distance, $x<\xi$, string effects are not pronounced and the wave function decays according to conventional WKB. At $\xi<x$ the action (\ref{9}) is proportional to $\ln(\gamma\tau)$ which again results in proportionality of the
action to $x$. The coefficient 1/2 in the action will be discussed below. We emphasize that the wave function (\ref{10}) is taken on the classical trajectory where each $x$ defines a full set
of string coordinates.

The curves 2 and 3 in Fig.~\ref{fig2} denote the points under the barrier where the wave function is exponentially small according to Eq.~(\ref{10}). The curve 1 represents the point of the classically allowed region where the particle is located at $x=0$ corresponding to the $\delta$ well. The wave function at this point is of the order of unity. The configuration 1 in Fig.~\ref{fig2} is related to the bound state of the particle in a strongly quantum well (with the energy $E_{p}=0$) attached to the elastic string (with the energy $V$). The string displacement is of the order of $\xi$ which is not proportional to $\hbar$. This length is a product of two macroscopic parameters, the particle velocity $\sqrt{V/m}$ and the friction attenuation time $1/\gamma$.

At a region of $x$ away from a narrow potential well, the potential is a constant and, as follows from Eq.~(\ref{5}), the typical time scale is of the order of $1/\gamma$. On the other hand, the same time scale is $y/s$ where $y$ is the typical length of the string segment involved into the game. It follows that $y\sim s/\gamma\sim m/\rho$. Even when coupling to the string is weak (a small mass density of the string $\rho$) the effect on the particle is not small since a long string segment is involved.

The sequence of the curves in Fig.~\ref{fig2} reminds a bow string which gives up its energy to the particle (``arrow''). This allows to call the joint state of particle and string as the quantum bow. Note that the displacement in bow state is macroscopic, that is not proportional to $\hbar$.
\subsection{Particle energy under the barrier}
The total energy $E$ on the trajectory does not depend on $\tau$ and consists of two contributions which separately depend on $\tau$. The first one is the particle energy
\begin{equation}
\label{7c}
E_{p}=-\frac{m}{2}\left(\frac{\partial x}{\partial\tau}\right)^{2}+V,
\end{equation}
which turns to zero at $\tau=0$. The level with zero energy corresponds to a free particle when $\rho=0$. The second part is the string energy, $E-E_{p}$. According to Eq.~(\ref{6}), the string stops at $\tau=0$ excepting the point $y=0$ which does not contribute to the string energy. This means, that the string energy at $\tau=0$ is determined solely by its elastic part which is, therefore, the total energy
\begin{equation}
\label{8}
E=\frac{\rho s^{2}}{2}\int^{\infty}_{-\infty}dy\left[\frac{\partial u(y,0)}{\partial y}\right]^{2}.
\end{equation}
It is easy to insert the expression (\ref{6}) into Eq.~(\ref{8}) and to perform the integrations. We drop simple calculations and the result is that the total energy of the system is $E=V$.

When $V(x)$ in Eq.~(\ref{2}) is not a constant it is more convenient, using Eq.~(\ref{7a}), to rewrite Eq.~(\ref{8}) in the form
\begin{equation}
\label{8a}
E=\frac{m\gamma}{\pi}\int^{\infty}_{0}d\tau_{1}\frac{\partial x(\tau_{1})}{\partial\tau_{1}}\int^{\infty}_{0}\frac{\partial x(\tau_{2})}{\partial\tau_{2}}\,\frac{d\tau_{2}}{\tau_{1}+\tau_{2}},
\end{equation}
where $x(\tau)$ satisfies Eq.~(\ref{7b}). The derivative can be estimated as $\partial x/\partial\tau\sim\sqrt{V/m}$. When in Eq.~(\ref{7b}) the force $V'(x)$ is moderate, a typical time is
of the order of $1/\gamma$. This means that even for a small damping coefficient $\gamma$, as follows from Eq.~(\ref{8a}), $E\sim V$.

The harmonic reservoir pushes the energy up to the barrier top. The enhancement of the ground state energy ($E=V$ instead of $E=0$ at $\rho=0$) is connected with a finite elastic energy of the string when $\partial u/\partial y$ is finite at the particle position, curve 1 in Fig.~\ref{fig2}.

\begin{figure}
\includegraphics[width=6.5cm]{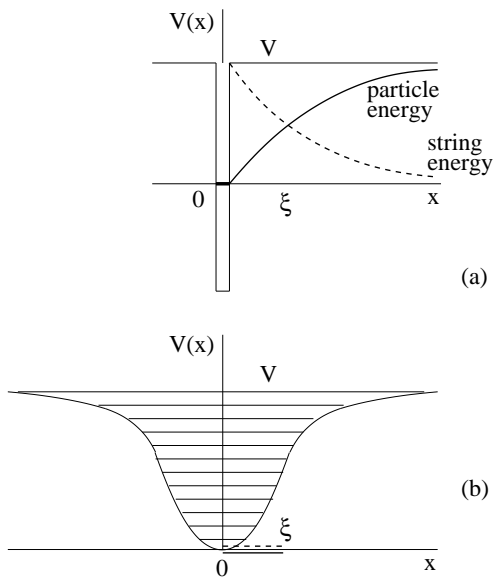}
\caption{\label{fig3}(a) Very quantum well. A particle energy (solid curve) and a string energy (dashed curve) are plotted along the classical trajectory under the barrier. Each point of the trajectory is marked by the coordinate $x$ which determines all other string coordinates along the trajectory. (b) Semiclassical well when level separation is less than $\hbar\gamma$. The
total energy (particle plus string) is zero within the semiclassical accuracy.}
\end{figure}

Particle and string energies are plotted in Fig.~\ref{fig3}(a) along the trajectory $u_{i}(\tau)$. The parameter $\tau$ parametrizes the trajectory. According to that, each value of $u_{0}(\tau)=x(\tau)$ (\ref{7}) determines the full set of other coordinates $u_{i}(\tau)$. For this reason, $x$ can parametrize the trajectory which is used in Fig.~\ref{fig3}(a).

One can calculate a contribution of the particle energy $E_{p}$ to the total energy $E=V$ along the classical trajectory as a function of $\tau$. Then one should substitute $\tau$ as a function of $x$ from Eq.~(\ref{7}). We obtain at small $x\ll\xi$ that the particle energy is small, $E_{p}\simeq 0$, and the string is most tensed (curve 1 in Fig.~\ref{fig2}). At a large $x$ the string becomes almost straight (curve 3 in Fig.~\ref{fig2}) and the particle energy is close to $V$
\begin{equation}
\label{11}
E_{p}=V\left[1-\frac{4}{\pi^{2}}\exp\left(-\frac{|x|}{\xi}\right)\right],\hspace{0.5cm}\xi\ll |x|.
\end{equation}
Since the particle energy is close to the barrier top propagation under the barrier becomes easier. This results in the reduced action (compared to WKB one) at $\xi\ll x$ in Eq.~(\ref{10}). The conclusion about increasing particle energy, as in Fig.~\ref{fig3}(a), can be drawn also from Eqs.~(\ref{7b}) and (\ref{4a}).

We calculated the ground state energy considering the wave function on the certain path in the multi-dimensional space. If to put $V(x)=V$ in Eq.~(\ref{2}) and to omit the $\delta$ function the potential becomes harmonic in all space and, after a diagonalization of the quadratic form, variables of the infinite set are separated. The wave function becomes a superposition of eigenfunctions of partial harmonic oscillators. Now to account for the $\delta$ function in Eq.~(\ref{2}) one has to write a wave function in the parts of the space with $x<0$ and with $0<x$ and then to match the both function at $x=0$ according to the jump of the derivative at the $\delta$ function. The total energy is infinite and is a sum of eigenvalues of partial harmonic oscillators. To obtain the finite energy $E=V$ one should subtract another infinite energy calculated without the $\delta$ function. This is a quantum mechanism of creation of a macroscopic displacement of a string (bow).

One can see that the bow state substantially differs from a polaronic state in a crystal. In a strongly coupled polaron an electron induces, by a classical polarization, a macroscopic lattice displacement which serves as a well to further reduce an electron (and total) energy. In contrast, in the bow state an electron does not reduce its energy since it is strongly coupled to the external well. For this reason, due to a connection to the environment the polaron energy is reduced but the bow energy is increased.
\subsection{Conditions for the bow state formation}
A formation of the bow state, which accompanies a particle in a well, is possible solely if the well is sufficiently quantum. When the potential well is not a $\delta$ function but a smooth well of the magnitude $V$ and of the width $a$ then the bow state exists under the condition $a<\xi$. This condition provides a minor influence of potential forces on the bow formation which occurs at the scale $\xi$. A level separation in the well can be estimated as $\hbar\Omega\sim (\hbar/a)\sqrt{V/m}$ where $\Omega$ is a classical oscillation frequency in the well. The above condition is equivalent to $\gamma<\Omega$. In this case the string shape has a form as in Fig.~\ref{fig2} which pulls the particle out of the well increasing its kinetic energy.

In contrast to that, for a semiclassical well, as in Fig.~\ref{fig3}(b), distance between discrete energy levels is relatively small, $\hbar\Omega<\hbar\gamma$. Under this condition the inequality $\xi< a$ holds and the bow could be formed close to the bottom of the well where the potential is harmonic. But it does not occur since the total energy differs from zero in Fig.~\ref{fig3}(b) due to harmonic quantum fluctuations only. This energy is proportional to $\hbar$ and is zero in the semiclassical approach. This is schematically shown in Fig.~\ref{fig3}(b) where a string (the dashed line) and a particle (the solid line) energies are close to zero.

For a quantum well as in Fig.~\ref{fig3}(a) the string configuration relates to Fig.~\ref{fig2} when the string pulls out the particle from the well. This increases a particle energy when it moves under the barrier. In contrast to that, in tunneling through a barrier formed, for example, by a cubic potential \cite{LEGGETT} the bow state is not formed due to a large potential force under the barrier. In this case the particle looses its energy under the barrier and the string configuration, of the type as in Fig.~\ref{fig1}, brakes the particle tending to return it to the well.

It is obvious from Eq.~(\ref{7b}) that a finite potential force $F=-V'(x)$ at the barrier region destroys the bow state. In this case the time scale becomes not large but determined by the Newtonian mechanics with no participation of the small friction term in Eq.~(\ref{7b}). In other words, for the bow formation the barrier should be flat at $x>\xi$. To approximately estimate a destructive role of a potential force $F$ one can use Eq.~(\ref{7b}). Since $\partial x/\partial\tau\sim\sqrt{V/m}$, the potential force in Eq.~(\ref{7b}) is small when
\begin{equation}
\label{11aa}
F<\frac{V}{\xi},
\end{equation}
which is a condition of the bow formation. We accounted for the definition (\ref{6a}).

We check the criterion (\ref{11aa}) in the case of a homogeneous electric field ${\cal E}_{0}$ acting on the particle. In this case the particle state in the well becomes metastable due to tunneling through the barrier but it is still characterized by a bow energy since the tunneling probability is small. One can show that
\begin{equation}
\label{11aaa}
{\rm bow}\hspace{0.12cm}{\rm energy}\sim V
\begin{cases}
1;& e{\cal E}_{0}<V/\xi\\
V/e{\cal E}_{0}\xi;& V/\xi<e{\cal E}_{0}
\end{cases}
\end{equation}
At a large electric field, when the condition (\ref{11aa}) does not hold, the bow energy tends to zero. A typical time $\tau_{0}$ of the process is $1/\gamma$ at a small electric field. At a large field it follows that $\tau_{0}\sim (V/e{\cal E}_{0}\xi)/\gamma$. The applicability of the semiclassical method, $V\tau_{0}>\hbar$, is violated at the large electric field
$e{\cal E}_{0}\simeq F_{c}$ where
\begin{equation}
\label{11aaaa}
F_{c}=\frac{V}{\xi_{q}}.
\end{equation}
Here $\xi_{q}=\sqrt{\hbar/m\gamma}$ is the quantum limit of the coherence length. The bow state energy tends to zero when the electric field reaches its large quantum limit.

Potential forces at a far distance (of the order of $\xi$) under the barrier, where the wave function is exponentially small, can essentially influence a classically allowed region where the wave function is not small. This occurs due to quantum coherence of the state extended over the distance of $\xi$.
\section{INTERACTION OF AN ATOM WITH A DIELECTRIC}
\label{atom}
In this section we analyze the certain interactions of an atom with a dielectric.
\subsection{Van der Waals interaction and the bow phenomena}
\begin{figure}
\includegraphics[width=6.5cm]{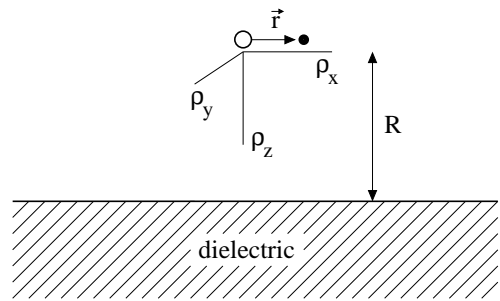}
\caption{\label{fig4}Hydrogen atom is separated by the distance $R$ from the dielectric. The nucleus (open circle) and the electron (dot) are connected by the vector $\vec r$.}
\end{figure}

When an atom is separated by the distance $R$ from a dielectric surface, as in Fig.~\ref{fig4}, an energy of electromagnetic zero point fluctuations depends on $R$ and this results in a force called van der Waals one. In the situation of Fig.~\ref{fig4} the van der Waals interaction has the form \cite{LANDAU1}
\begin{equation}
\label{11a}
U_{vdW}(R)=-\frac{3\hbar c\alpha}{8\pi R^{4}}\,\frac{\varepsilon_{0}-1}{\varepsilon_{0}+1}\,\varphi(\varepsilon_{0}),
\end{equation}
where $\varepsilon_{0}$ is a static permittivity of the dielectric and $\alpha$ is a static polarizability of the atom. Eq.~(\ref{11a}) is valid at the distance $R>c/\Omega_{0}\sim 500\AA$ where
$\Omega_{0}$ is an atomic frequency.

The function $\varphi(\varepsilon_{0})$ is of the order of unity and is determined in Ref.~\cite{LANDAU1}. For the ground state of the hydrogen atom $\alpha=9a^{3}_{B}/2$ where $a_{B}=\hbar^{2}/me^{2}$ is the Bohr radius \cite{LANDAU}. Taking, for example, $\varepsilon_{0}=2$ and accounting for the relation $\hbar c/e^{2}\simeq 137$ one can write Eq.~(\ref{11a}) in the form
\begin{equation}
\label{11b}
U_{vdW}(R)\simeq-37.3\,V\left(\frac{a_{B}}{R}\right)^{4},
\end{equation}
where $V=me^{4}/2\hbar^{2}$ is one rydberg.

The van der Waals interaction (\ref{11a}) is proportional to $\hbar$ since it has a fluctuation (zero mean field) origin. Virtual transitions between levels in the atom (see Fig.~\ref{fig0}) produce a fluctuating electromagnetic field. Fluctuations of the atomic electron occur inside a small region of the size of the Bohr radius.

But apart from these short range fluctuation phenomena there are macroscopic (nonzero mean field) ones occurring at the large scale $\xi$ (\ref{6a}) as described in Sec.~\ref{bow}. This is a formation of the bow state at that scale. A role of a string displacement (as in Fig.~\ref{fig2}) is played by a potential of the electric field.

To go from those general features to more detailed description let us consider the atom to be on the distance $R$ from the dielectric surface as shown in Fig.~\ref{fig4}. To study a wave function of the system one can track it along some classical trajectory in imaginary time $t=-i\tau$ in the multi-dimensional space of $\vec r$ and an infinite set of electromagnetic coordinates.

When the electron is sufficiently far from the nucleus the Coulomb force is small. However a classical motion in imaginary time is not free since the moving electron produces a nonstationary electric field in the dielectric. This field results in energy dissipation due to an imaginary part of the dielectric permittivity. Therefore a classical equation of motion in imaginary time should be of the same type as Eq.~(\ref{5}) where the second term is due to a dissipation.
\subsection{Classical dissipation of an atomic electron in the presence of a dielectric}
If the electron moves with a positive energy (classically) in the Coulomb field of the nucleus one can determine its nonstationary field and to calculate, due to dissipation in the dielectric, a damping coefficient $\gamma$ in the classical equation (\ref{5}).

For simplicity we consider the hydrogen atom. The total electric field is a sum of the proton and the electron contributions (see Fig.~\ref{fig4})
\begin{equation}
\label{12}
\vec{\cal E}(\vec\rho,t)=\vec{\cal E}_{pr}(\vec\rho)+\vec{\cal E}_{el}[\vec\rho-\vec r(t)],
\end{equation}
The energy dissipation in the dielectric is \cite{LANDAU2}
\begin{equation}
\label{13}
\frac{dE}{dt}=\frac{1}{4\pi}\int\vec{\cal E}(\vec\rho,t)\frac{\partial\vec{\cal D}(\vec\rho,t)}{\partial t}\,d^{3}\rho.
\end{equation}
In Eq.~(\ref{13}) $E$ is an energy of the electric field and $\vec{\cal D}$ is an electric displacement. In Fourier components
$\vec{\cal D}_{\omega}(\vec\rho)=\varepsilon(\omega)\vec{\cal E}_{\omega}(\vec\rho)$, where $\varepsilon(\omega)=\varepsilon'(\omega)+i\varepsilon''(\omega)$, and the integration is performed inside the dielectric, see Fig.~\ref{fig4}. We consider an isotropic
permittivity tensor $\varepsilon_{ik}=\varepsilon\delta_{ik}$.

The term in Eq.~(\ref{13}) with squared velocity is produced by the part $\vec{\cal E}_{el}\partial\vec{\cal D}_{el}/\partial t$ of Eq.~(\ref{13}) since the term
$\vec{\cal E}_{pr}\partial\vec{\cal D}_{el}/\partial t$ is a full time derivative and does not contributes to dissipation. The proper part is
\begin{eqnarray}
\nonumber
&&\frac{dE}{dt}=\frac{1}{4\pi}\int d^{3}\rho\,\vec{\cal E}_{el}[\vec\rho-\vec r(t)]\int^{\infty}_{-\infty}dt_{1}\vec{\cal E}_{el}[\vec\rho-\vec r(t+t_{1})]\\
\label{14}
&&\int^{\infty}_{-\infty}\frac{d\omega}{2\pi}(-i\omega)\varepsilon(\omega)\exp(i\omega t_{1}).
\end{eqnarray}
According to analytical properties of $\varepsilon(\omega)$ (causality), the $\omega$ integral in Eq.~(\ref{14}) equals zero at negative $t_{1}$ \cite{LANDAU2}. The argument of the second electric field in Eq.~(\ref{14}) can be written as $\vec\rho-\vec r(t)-t_{1}d\vec r/dt$ since at a large distance from the nucleus (only that distance is relevant) the electron velocity is a constant.

We track the wave function $\psi(\vec r)$ of the electron along the direction $x$. Note that $\vec r=\{x,y,z\}$. Nonzero value of Eq.~(\ref{14}) corresponds to the second order in $dx/dt$ and higher terms
\begin{eqnarray}
\nonumber
\label{15a}
&&\frac{dE}{dt}=-\frac{V}{4(1+\varepsilon_{0})^{2}}\frac{a_{B}}{R^{3}}\bigg[\left(\frac{dx}{dt}\right)^{2}\frac{\partial\varepsilon^{''}(\omega)}{\partial\omega}\bigg|_{\omega=0}\\
&&+\frac{C}{R^{2}}\left(\frac{dx}{dt}\right)^{4}\frac{\partial^{3}\varepsilon^{''}(\omega)}{\partial\omega^{3}}\bigg|_{\omega=0}+...\bigg],
\end{eqnarray}
where $C$ is a number of the order of unity. Since in Eq.~(\ref{15a}) the low frequency limit is relevant we use the static expression for the electric field in the dielectric \cite{LANDAU2}
\begin{equation}
\label{16}
\vec{\cal E}_{el}(\vec\rho)=-\frac{2}{1+\varepsilon_{0}}\,\frac{|e|\vec\rho}{\rho^{3}}.
\end{equation}

The classical equation of motion has the form
\begin{equation}
\label{15}
m\frac{\partial^{2}x}{\partial t^2}+\frac{1}{\partial x/\partial t}\frac{dE}{dt}+V'(x)=0,
\end{equation}
which corresponds to the energy dissipation by the particle
\begin{equation}
\label{15aa}
\frac{d}{dt}\left[\frac{m}{2}\left(\frac{\partial x}{\partial t}\right)^{2}+V(x)\right]=-\frac{dE}{dt}.
\end{equation}
\subsubsection{Large $R$ (ohmic dissipation)}
For sufficiently large $R$ the first term in the series (\ref{15a}) dominates and the dissipation becomes ohmic. It follows from comparison of Eqs.~(\ref{5}) and (\ref{15}) that
\begin{equation}
\label{17}
\frac{\hbar\gamma}{V}=\frac{V}{\hbar\omega_{0}}\left(\frac{a_{B}}{R}\right)^{3},
\end{equation}
where
\begin{equation}
\label{18}
\frac{1}{\omega_{0}}=\frac{1}{2(1+\varepsilon_{0})^{2}}\frac{\partial\varepsilon^{''}(\omega)}{\partial\omega}\bigg|_{\omega=0}.
\end{equation}
To account for a frequency dispersion of $\varepsilon(\omega)$ one can use Debye's formula relevant for a relaxation of dipole moments in a dielectric \cite{FRO}
\begin{equation}
\label{18a}
\varepsilon(\omega)=\varepsilon_{\infty}+\frac{\varepsilon_{0}-\varepsilon_{\infty}}{1-i\omega/\omega_{0}}.
\end{equation}
This leads to the estimate
\begin{equation}
\label{19}
\frac{\hbar\gamma}{V}\simeq\frac{\varepsilon_{0}-\varepsilon_{\infty}}{2(1+\varepsilon_{0})^{2}}\,\frac{V}{\hbar\omega_{0}}\left(\frac{a_{B}}{R}\right)^{3}.
\end{equation}

Domination of the first term in Eq.~(\ref{15a}) is possible when the parameter $(t_{1}/R)dx/dt$ in Eq.~(\ref{14}) is small. Since $t_{1}\sim 1/\omega_{0}$ and in the underbarrier problem a typical velocity is $\sqrt{V/m}$, Eq.~(\ref{15}) is valid when
\begin{equation}
\label{20}
a_{B}\frac{V}{\hbar\omega_{0}}<R\hspace{0.5cm}({\rm large}\,R).
\end{equation}
Here $V$ is one rydberg, $a_{B}$ is the Bohr radius, and $\omega_{0}\simeq 10^{10}\,{\rm s}^{-1}$ for usual dipole dielectrics \cite{FRO}. Since $V\simeq 10^{16}\,{\rm s}^{-1}$ the condition (\ref{20}) reads  $10^{6}a_{B}<R$. This determines the limit of the large $R$.
\subsubsection{Intermediate $R$ (nonohmic dissipation)}
When the condition (\ref{20}) of a large $R$ does not hold the entire series (\ref{15a}) is essential. In this case it is more convenient to use the limit of a large $(t_{1}/R)dx/dt$ in Eq.~(\ref{14}) when the $\rho_{z}$ integration can be started from zero. After a not long calculation we arrive to
\begin{equation}
\label{19a}
\frac{dE}{dt}=\frac{e^{2}}{2\pi|\partial x/\partial t|}\int^{\infty}_{0}d\omega\omega\varepsilon''(\omega)\int^{\infty}_{t_{0}}\frac{dt_{1}}{t_{1}}\cos\omega t_{1},
\end{equation}
where $t_{0}\simeq R/(\partial x/\partial t)$. Evaluating the time integration, one can represent Eq.~(\ref{19a}) in the form
\begin{equation}
\label{19aa}
\frac{dE}{dt}\simeq\frac{e^{2}}{2\pi|\partial x/\partial t|}\int^{\Omega}_{0}d\omega\omega\varepsilon''(\omega)\ln\left(\frac{1}{\omega R}\,\frac{\partial x}{\partial t}\right),
\end{equation}
where $\Omega\simeq \left(1/R\right)\partial x/\partial t$.

A result of the integration in Eq.~(\ref{19aa}) depends on frequency dispersion of the permittivity. For a polar dielectric, besides the low frequency region $\omega\sim\omega_{0}$ connected to 
the dipole relaxation (\ref{18a}), there are the phonon region, $\omega\sim\omega_{ph}=10^{13}\,{\rm s}^{-1}$, and the atomic region, $\omega\sim V/\hbar\simeq 2\times 10^{16}\,{\rm s}^{-1}$ \cite{FRO}. We specify the interval of intermediate $R$ as
\begin{equation}
\label{21}
a_{B}\frac{V}{\hbar\omega_{ph}}<R<a_{B}\frac{V}{\hbar\omega_{0}}\hspace{0.5cm}({\rm intermediate}\,R).
\end{equation}
Under this condition the integral in Eq.~(\ref{19aa}) can be estimated as $\Omega^{2}$ since $\Omega<\omega_{ph}$. Substituting this estimate into Eqs.~(\ref{19aa}) and (\ref{15}) we obtain the classical equation of motion valid under the condition (\ref{21})
\begin{equation}
\label{21aa}
m\frac{\partial^{2}x}{\partial t^2}+\frac{e^{2}}{R^{2}}\,{\rm sgn}\left(\frac{\partial x}{\partial t}\right)+V'(x)=0.
\end{equation}
The second term relates to a substantially nonohmic dissipation. When the dissipation $dE/dt$ is not quadratic with respect to velocity we call that dissipation nonohmic.
\subsubsection{Small $R$ (nonohmic dissipation)}
The small region of $R$ is specified as
\begin{equation}
\label{211}
a_{B}<R<a_{B}\frac{V}{\hbar\omega_{ph}}\hspace{0.5cm}({\rm small}\,R).
\end{equation}
Under this condition the upper limit of integration in Eq.~(\ref{19aa}) is in the interval $\omega_{ph}<\Omega<V/\hbar$. This interval lies higher than a region of phonon absorption peaks and therefore $\varepsilon''(\omega)$ is small at that interval. For this reason, the whole phonon part of frequencies mainly contribute to the integral in Eq.~(\ref{19aa}) which can be estimated
as $\omega^{2}_{ph}$. Substituting this estimate into Eqs.~(\ref{19aa}) and (\ref{15}) we obtain the classical equation of motion valid under the condition (\ref{211})
\begin{equation}
\label{21a}
m\frac{\partial^{2}x}{\partial t^2}+\frac{e^{2}\omega^{2}_{ph}}{\partial x/\partial t\,|\partial x/\partial t|}+V'(x)=0.
\end{equation}
The second term relates to a nonohmic dissipation. In the limit considered the velocity $\partial x/\partial t$ cannot be very small.
\section {OHMIC DISSIPATION}
\label{ohm}
In this section we consider the case of ohmic dissipation related to the condition (\ref{20}).

Since the dissipation is determined by a squared velocity, the classical equation of motion has the form (\ref{5}) with the damping coefficient (\ref{17}). Eq.~(\ref{5}) can be obtained from Eq.~(\ref{7b}) if to substitute the integration in $\tau_{1}$ by a continuous contour plus the pole part at the point $\tau$. After that one can continue imaginary time to the real axis and to get Eq.~(\ref{5}).

But the inverse procedure, obtaining Eq.~(\ref{7b}) from Eq.~(\ref{5}), is correct when a reservoir is a harmonic string as in Sec.~\ref{bow}. Generally, one can be an additional factor, of the
type $\cos k[x(\tau)-x(\tau_{1})]$, under the integral in Eq.~(\ref{7b}). In the case of Josephson junctions $k$ is a constant but in the polaronic problem in crystals there is an integration with respect to the wave vector $k$ \cite{WEISS}. That factor does not influence the derivation of the classical equation (\ref{5}) as it turns to unity at the pole $\tau_{1}=\tau$.

In our problem the electron is outside a crystal at a distance much larger than a period of a crystal lattice. The electron interacts with electromagnetic waves only and the typical wave vector
is $k\sim\gamma/c$. Since a typical distance is $x\sim\xi$ one can obtain the estimate $kx\sim e^{2}/\hbar c\simeq 1/137$. For this reason, the above cosine factor is close to unity and we arrive to Eq.~(\ref{7b}) for $x(\tau)$. At $a_{B}<r<\xi$ the electron wave function is proportional to $\exp(-r\sqrt{2mV}/\hbar)$ which serves as a condition (at a small $r$) for the region $r\sim\xi$. This is analogous to the condition (\ref{4a}) to Eq.~(\ref{7b}).

In other words, Eq.~(\ref{7b}), valid for a harmonic reservoir, holds also in our case for the trajectory in the whole space projected on the electron coordinates $\vec r=\{x(\tau),0,0\}$. This result is valid for any trajectory parallel to the dielectric surface in Fig.~\ref{fig4}. This means that in the entire plane the wave function depends on $|\vec r|$ where $\vec r=\{x,y,0\}$. Properties of the quantum state do not depend on that along which trajectory we probe it. A choice of a trajectory with a finite $z$ component is less convenient since it will be a coordinate dependence of the friction in that case and mapping on the harmonic string is not convenient.

The coherence length (\ref{6a}) can be written in the form
\begin{equation}
\label{20aa}
\xi=a_{B}\frac{2V}{\pi\hbar\gamma}.
\end{equation}
Since the bow state is macroscopic it can exist under the semiclassical condition $\hbar\gamma<V$. It follows the estimate in the ohmic case
\begin{equation}
\label{20a}
\xi\simeq a_{B}\frac{\hbar\omega_{0}}{V}\left(\frac{R}{a_{B}}\right)^{3}\hspace{0.5cm}({\rm large}\,R).
\end{equation}
\subsection {NONOHMIC DISSIPATION}
\label{nohm}
In this section we consider the case of nonohmic dissipation related to the conditions (\ref{21}) and (\ref{211}).

Whereas in our ohmic case the classical dissipation (\ref{5}) results in Eq.~(\ref{7b}) for imaginary time, in the nonohmic case it is not clear what is an analogue of the classical equation (\ref{21a}) in imaginary time. However one can do an estimate for the coherence length in the nonohmic case. If to put in the dissipative term of Eq.~(\ref{21a})
$\partial x/\partial t\sim i\sqrt{V/m}$ one can approximately write
\begin{equation}
\label{201}
m\left(\frac{\partial x}{\partial\tau}-\sqrt{\frac{2V}{m}}\right)\sim\tau\,\frac{e^{2}}{R^{2}}.
\end{equation}
The characteristic time $\tau_{0}$ of the trajectory is determined by the condition that a change of the velocity $\partial x/\partial\tau$ is of the order of $\sqrt{V/m}$. This gives
$\tau_{0}\sim(\hbar/V)(R/a_{B})^{2}$. The coherence length can be estimated as $\xi\sim\tau_{0}\,\partial x/\partial\tau$ which results in the expression
\begin{equation}
\label{202}
\xi\sim a_{B}\left(\frac{R}{a_{B}}\right)^{2}\hspace{0.5cm}({\rm intermediate}\,R).
\end{equation}
The same arguments are applicable to the case of small $R$ (\ref{211}). Analogously we obtain $1/\tau_{0}\sim (V/\hbar)(\hbar\omega_{ph}/V)^{2}$, which plays a role of a classical damping coefficient, and the
coherence length
\begin{equation}
\label{203}
\xi\sim a_{B}\left(\frac{V}{\hbar\omega_{ph}}\right)^{2}\hspace{0.5cm}({\rm small}\,R).
\end{equation}
\begin{figure}
\includegraphics[width=7cm]{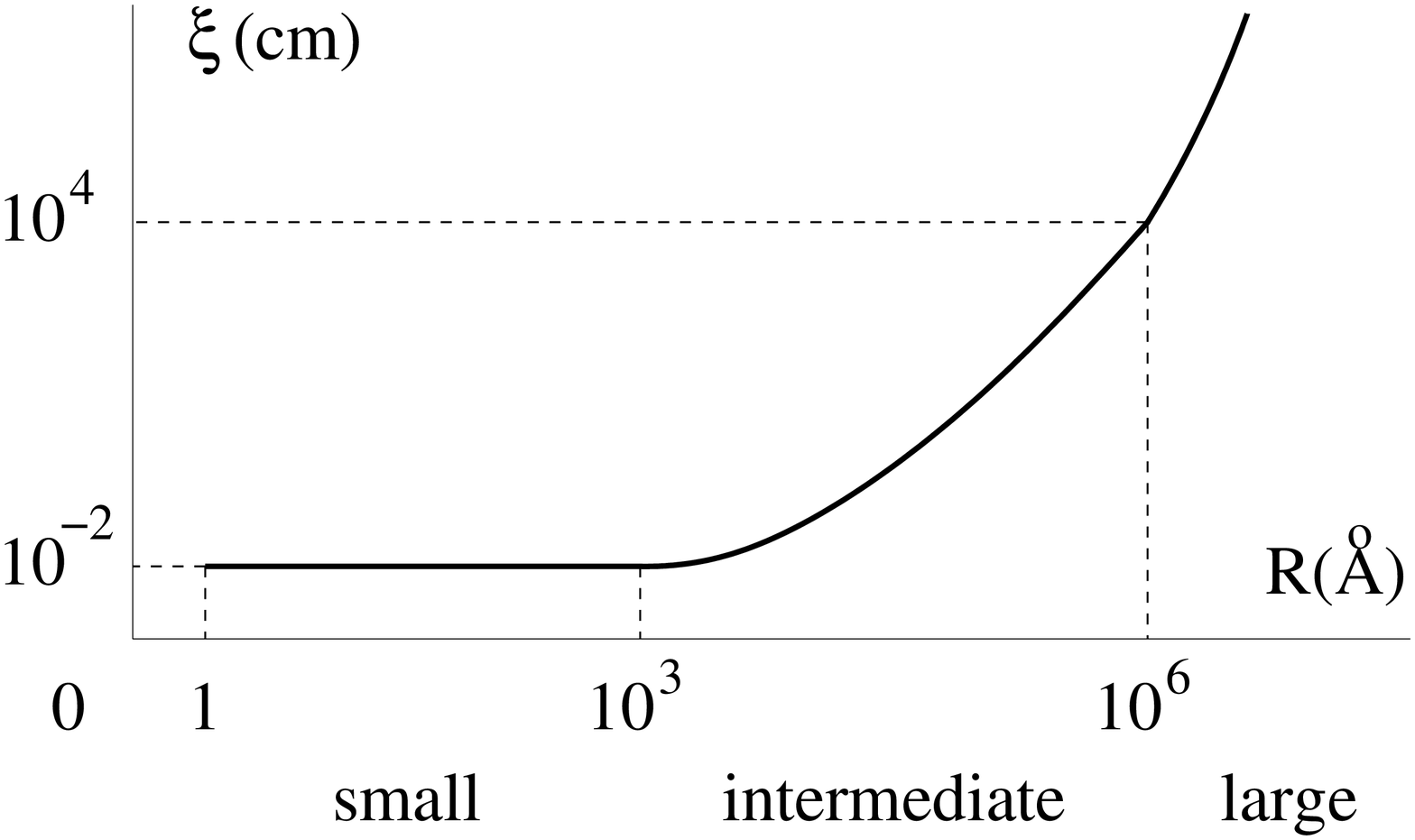}
\caption{\label{fig5}Coherence length as a function of the distance $R$ between the atom and the dielectric as in Fig.~\ref{fig4}. See the text for details.}
\end{figure}

The results (\ref{20}), (\ref{21}), (\ref{211}), and (\ref{20a}), (\ref{202}), (\ref{203}) are summarized in Fig.~\ref{fig5} where the parameters $V/\hbar\omega_{0}=10^{6}$, $V/\hbar\omega_{ph}=10^{3}$, and $a_{B}\simeq 1\,\AA$ are taken.
\section{BOW STATE OF THE ATOM}
\label{atom1}
In the method used we track the electron wave function $\psi(\vec r)$ along the classical trajectory $\vec r=\{x(\tau),0,0\}$ and $x(\tau)$ obeys a classical equation of motion in imaginary time. According to this equation, along the trajectory the electron gains the energy $V$ stored by the reservoir as in Fig.~\ref{fig3}(a). This means that the ground state of the system "atom and reservoir" is the bow state whose energy is higher than one of a noninteracting atom by $V$. The electron wave function along the direction $\vec r=\{x,0,0\}$ is of the type of one given by Eq.~(\ref{10}) when $x$ is not too close to the nucleus. Tracking the wave function along a direction with a finite $z$ component (Fig.~\ref{fig4}) is not convenient. In this case the dissipation will depend on coordinate as the problem is not homogeneous in $z$.

At a finite temperature a role of a potential is played by the free energy which also depends on $R$. Since the ground state (of the order of rydberg) is much larger than $T$, the free energy will be mainly the ground state. So the above results are valid at room temperature.

In the absence of a dielectric, the electric field of the hydrogen atom in the ground state decreases exponentially at the distance $a_{B}$ \cite{LANDAU}. The energy transfer along the classical trajectory in imaginary time is similar to one between the particle and the string in Fig.~\ref{fig2}. The electron energy has its maximal value (minimal energy of the reservoir) far from the atom ($\tau=\infty$) where the electron wave function is exponentially small. With the reduction of $\tau$ the electron approaches the atom and its energy reduces due to transfer into the reservoir. At $\tau=0$ (the atom position) the electron energy takes its minimal value, as in the bare atom, and an energy of the reservoir reaches the maximum, $V$. These energies correspond to the quantum state (bow) of the system "atom above dielectric" since at the trajectory point $\tau=0$ an electron wave function is not small.

A role of the reservoir is played by the electric field associated with polarizations of the atom and the dielectric. In the ground state of the whole system the mean (nonfluctuating) electric field $\vec{\cal E}_{bow}(\vec r)$ is created. One can estimate its value from the balance between the electrostatic energy in the volume $\xi^{3}$, that is ${\cal E}^{2}_{bow}\xi^{3}$, and the energy $V$. This condition results in the estimate $e{\cal E}_{bow}\sim(V/\xi)\sqrt{\hbar\gamma/V}$. So, in Eq.~(\ref{7b}) the potential force $e{\cal E}_{bow}$ is small according to the condition (\ref{11aa}). In contrast to the harmonic string considered above, it is not easy to determine all details of $\vec{\cal E}_{bow}$. One can say that this field has an axial symmetry (for an isotropic permittivity tensor $\varepsilon_{ik}=\varepsilon\delta_{ik}$) and is originated from a charge redistribution occurring at the distance $\xi$ inside the dielectric. This is some sort of a ferroelectric state but of a completely different origin than the usual ferroelectricity.
\begin{figure}
\includegraphics[width=7cm]{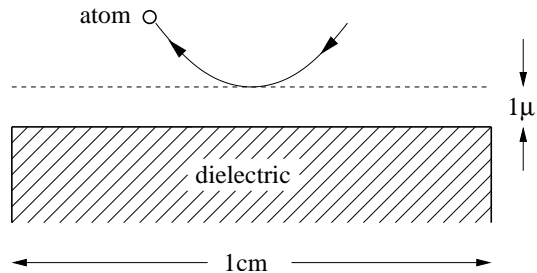}
\caption{\label{fig6}The bow state of the atom is formed when it is close to the surface of the dielectric, within the layer of $1\mu$ width. The moving atom is reflected from that layer to 
avoid the energy increase due to the bow formation.}
\end{figure}

In principle, there is a possibility of generation of some additional electric field to destroy bow state, according to the condition (\ref{11aa}), and to reduce the ground state energy by $V$. This can occur when the generated electric field is of the order of $V/e\xi$ in the volume of $\xi^{3}$. The electrostatic energy pay is proportional to $V\xi/a_{B}$ which is much large than the energy gain $V$. For this reason, an electric field is not generated and the bow state of a single atom survives.

An unusual feature of the bow state of the system "atom above dielectric" is a totally quantum nature of the state with the coherence length $\xi$. The coherence length depends on the distance
$R$ between the atom and the dielectric as shown in Fig.~\ref{fig5}. For larger $R$ the bow state becomes more soft that is with a larger coherence length. It is clear that only regimes with
small and intermediate $R$ can be practically realized.

The criterion of the bow state, smallness of potential forces (\ref{11aa}), should be fulfilled at the distance $\xi$ on the dielectric surface for each fixed $R$. When the atom is far from the dielectric, the coherence length is large and the force (\ref{11aa}) parallel to the dielectric surface is small. The coherence length cannot be larger than linear sizes of the dielectric surface. For this reason, the bow state exists if the atom is close to the dielectric ($\xi$ is not large) and it is destroyed when the atom is far ($\xi$ is large). For the size of the surface of the dielectric of 1\,cm, as in Fig.~\ref{fig6}, the condition $\xi<1$\,cm is equivalent to $R<1\mu$. That is the bow state exists for an atom which is not far from the dielectric than $1\mu$. Since the bow state is of a larger energy (of the order of rydberg) the atom will be repelled from the region of the width of $1\mu$ near the dielectric surface. This unusual situation is shown in Fig.~\ref{fig6}.

A dielectric near the atom is essential for the phenomenon. In this case the classical dynamics of the electron corresponds to Eq.~(\ref{5}). In the absence of the dielectric (an isolated atom) the classical frictional force is due to wave radiation and the classical equation, instead Eq.~(\ref{5}), has the form \cite{LANDAU4}
\begin{equation}
\label{206}
m\frac{\partial^{2}x}{\partial t^2}-\frac{2e^{2}}{3c^{3}}\,\frac{\partial^{3}x}{\partial t^{3}}=-V'(x).
\end{equation}
As known, this equation has a meaning only when the damping force is smaller than a potential one. Therefore a semiclassical description (with a friction) of an electron of an isolated atom is impossible.

In this paper we consider a single atom above the dielectric. The situation, when the dielectric is in a gas of atoms, requires a special investigation. First, in the classical energy dissipation (\ref{14}) the electric field becomes a sum of contributions of various atoms and crossterms will modify the energy dissipation. Second, one has to take into account potential
forces acting on a moving electron from other atoms. In the gas of atoms the bow state is expected to be collective and this is to be studied.
\section{INTERPRETATION OF THE BOW STATE IN AN ATOM}
\label{interp}
According to Eq.~(\ref{10}), along the underbarrier classical trajectory a wave function of the electron remains of WKB type at distances $r$ less than the coherence length $\xi$. At larger distances the wave function is different. Tracking the wave function along the classical trajectory one can see that at $r>\xi$ the electron acquires the certain extra energy. Since the total
energy (electron + reservoir) is a constant along the trajectory, at $r<\xi$ the reservoir energy is higher by the same extra part. This is the bow state energy as in Fig.~\ref{fig3}(a). Therefore at large distances the electron wave function gets a fraction of high energy states.

There is a quantum mechanical analogy which enables to interpret the bow phenomenon studying just electron states. The atom and the dielectric can interact by fluctuating dipole
momenta \cite{LANDAU}
\begin{equation}
\label{205}
v(\vec r_{1},\vec r_{2})=\frac{\vec d_{1}\vec d_{2}}{r^{3}}-3\frac{(\vec d_{1}\vec r_{1})(\vec d_{2}\vec r_{2})}{r^{5}},
\end{equation}
where $r=|\vec r_{1}-\vec r_{2}|$, the atom 1 is above the surface, and the atom 2 belongs to the dielectric. A summation with respect to positions of the atom 2 should be done \cite {LANDAU1}. In the second order of the perturbation theory, with respect to $v$, the energy correction gives the van der Waals interaction \cite{LANDAU,LANDAU1}.

The same perturbation theory can be applied to calculations of the electron wave function of the ground state
\begin{equation}
\label{204}
\psi_{0}=\psi^{(0)}_{0}+\sum_{n}\frac{v_{n0}}{E_{0}-E_{n}}\,\psi^{(0)}_{n}+...
\end{equation}
where $\psi^{(0)}_{0}$ and $E_{n}$ pertain to unperturbed state of the system of the atom 1, above the surface, and the atom 2 of the dielectric \cite{LANDAU1}.

When states $n$ are of the continuous spectrum they do not exponentially decay with distance and the decay of $\psi_{0}$ is provided by interference of various oscillating contributions which mutually cancel each other. Every order is well convergent and the true wave function hardly differs from the unperturbed one. Indeed, at $x\sim\xi$ it is impossible to obtain within the perturbation theory the nonregular dependence $\psi\sim\psi^{(0)}\exp(V/\hbar\gamma)$ with respect to $\gamma$ following from Eq.~(\ref{10}).

Dissipation effects on the electron motion can violate in Eq.~(\ref{204}) the interference of various contributions from the continuous spectrum. Analogously, in disordered solids inelastic effects destroy electron interference leading to reduction of their localization \cite{KHM,GANT}. As a result of dephasing effects, the wave function is less decayed at large distances, as in Eq.~(\ref{10}), since it gets a fraction of overbarrier states which are not completely compensated.

In the classical language this is equivalent to acceleration of the electron by some mean electric field related to the bow state. This can be seen if to track the underbarrier wave function along the classical trajectory in imaginary time. In this process the electron gets more energy at large distances as in Fig.~\ref{fig3}(a).

Therefore the bow state is a result of violation of the interference of fluctuating states, Eq.~(\ref{204}), in the presence of the dissipation. In the absence of a dielectric the usual processes in quantum electrodynamics, with the classical analogue (\ref{206}), do not result in the bow state.
\section{ANOMALOUS LAMB SHIFT}
\label{lamb}
The Lamb shift $\delta E_{L}$ of atomic energy levels is due to the interaction of the atomic electron with fluctuating electromagnetic fields in vacuum \cite{LANDAU3}. The Lamb shift due to the interaction with fluctuations of a different nature, capillary waves on a surface of liquid helium, was considered in Ref.~\cite{DYK}.

The energy of the bow state, of the order of rydberg, is an energy of the environment. In the case of the system "atom above dielectric" it is due to a dielectric polarization. The typical scale of the problem $\xi$ is much larger than the Bohr radius $a_{B}$ where Lamb phenomena occur. Therefore the bow and the Lamb phenomena are well separated in space and can be considered independently.

As known, in the mechanism of the Lamb shift the typical frequency is higher that $V/\hbar$. Therefore the wavelength of relevant fluctuating fields is shorter that $a_{B}(\hbar c/e^{2})$ which, in turn, is much less that the spatial scales involved into the bow problem. These scales are determined by the geometry but not by an electromagnetic wave vector which is almost zero in the bow case. For this reason, one can consider two independent contributions to the total Lamb shift. One of them ($\delta E_{L}$) is due to the usual mechanism when a single atom interacts with electromagnetic fluctuations in the infinite space. The second one ($\delta E^{(D)}_{L}$) is related to electromagnetic fluctuations in the system "atom above dielectric". The upper index $D$ points out to the presence of a dielectric. The total Lamb shift $\delta E_{L}+\delta E^{(D)}_{L}$ is anomalous.

The energy of the atomic electron (\ref{7c}) at $\tau=0$ is unperturbed by the environment in the semiclassical approximation. But in some cases (for spectroscopy purposes, for example) it is worth to know a shift of the electron energy despite it is less compared to rydberg.

The usual Lamb shift $\delta E_{L}$ is due to an interaction of an electron in the atom with zero point electromagnetic fluctuations. The electron "vibrates" and smears out in the space. Its interaction with the nucleus is reduced and the electron level gets higher, $\delta E_{L}>0$. As known, it is impossible to calculate $\delta E_{L}$ within finite orders of the perturbation theory with respect to the coupling constant $g=e^{2}/\hbar c\simeq 1/137$ \cite{LANDAU3}. In the hydrogen atom the Lamb shift diverges at small photon energies $\omega$ as
\begin{equation}
\label{23}
\delta E_{L}\sim Vg^{3}\ln\frac{mc^{2}}{\hbar\omega},
\end{equation}
where $V$ is one rydberg. The infrared divergence (\ref{23}) should be cut off at the relatively small frequency of the Bohr scale, $\omega\sim V/\hbar$. After that the Lamb shift in the hydrogen atom takes the form
\begin{equation}
\label{24}
\delta E_{L}\sim Vg^{3}\ln\frac{1}{g}\,.
\end{equation}
As one can see from Eq.~(\ref{24}), $\delta E_{L}$ is not a regular function of the coupling constant $g$. This corresponds to an infinite order of the perturbation series despite smallness
of the semiclassical parameter $\delta E_{L}/\hbar\omega$. One can say that the Lamb shift phenomenon, which has a fluctuation nature, is between perturbation theory and semiclassical approach.

In our case of "atom above dielectric" there is an additional mechanism of the Lamb shift compared to the usual one. This mechanism is due to an interaction with the electromagnetic environment
in the presence of the dielectric.

We start with the case of the large distance between the atom and the dielectric (\ref{20}). In classical mechanics the energy dissipation of a moving electron is 
$dE/dt=-m\gamma(\partial\vec r/\partial t)^{2}$. In the quantum case this corresponds to a transition probability per unit time between the states $s$ and $s'$ \cite{LANDAU3}
\begin{equation}
\label{25}
w_{ss'}=\frac{m\gamma}{\hbar^{2}}(E_{s}-E_{s'})|\vec r_{ss'}|^{2},
\end{equation}
where $(E_{s}-E_{s'})$ is positive. The imaginary part of the energy, ${\rm Im}E_{s}=-\hbar/2\sum_{s'} w_{ss'}$, has the form
\begin{equation}
\label{26}
{\rm Im}E_{s}=-\frac{m\gamma}{2\hbar}\sum_{s'}(E_{s}-E_{s'})|\vec r_{ss'}|^{2}
\end{equation}
or equivalently
\begin{equation}
\label{27}
{\rm Im}E_{s}=-\frac{m\gamma}{2}\int^{\infty}_{0}d\omega\sum_{s'}(E_{s}-E_{s'})|\vec r_{ss'}|^{2}\delta(E_{s}-E_{s'}-\hbar\omega),
\end{equation}
where all $s'$ participate in the summation. Eq.~(\ref{26}) is ultimately a consequence of the unitarity condition. Eq.~(\ref{27}) is an imaginary part of the equation for the energy correction \cite{LANDAU3}
\begin{equation}
\label{28}
\delta E_{s}=\frac{m\gamma}{2\pi}\int^{\infty}_{0}d\omega\sum_{s'}|\vec r_{ss'}|^{2}\frac{E_{s}-E_{s'}}{E_{s}-E_{s'}-\hbar\omega+i0}.
\end{equation}
We consider the ground state energy, $s=0$, which has a real part only. This part can be estimated from the logarithmically divergent integral which should be cut off by $mc^{2}$ at the upper limit. At the lower limit one can approximately put $V$. Within the logarithmic accuracy
\begin{equation}
\label{29}
\delta E^{(D)}_{L}=\frac{m\gamma}{2\pi\hbar}\ln\left(\frac{mc^{2}}{V}\right)\sum_{s'}|\vec r_{0s'}|^{2}(E_{0}-E_{s'}).
\end{equation}
In Eq.~(\ref{29}) for the shift of the ground state energy we use the notation $\delta E^{(D)}_{L}$. The sum in Eq.~(\ref{29}) equals to $\hbar^{2}/2m$ according to the known sum rule \cite{ZIM}. Finally we obtain for the Lamb shift
\begin{equation}
\label{30}
\delta E^{(D)}_{L}=\frac{\hbar\gamma}{2\pi}\ln\frac{\hbar c}{e^{2}}.
\end{equation}
The result (\ref{30}) has a simple meaning. The energy shift is mainly the classical damping coefficient multiplied by $\hbar$. For $\gamma$ given by Eqs.~(\ref{19}) and (\ref{20}) the energy shift (\ref{30}) is too small compared to the usual Lamb shift (\ref{24}).

In contrast, in the limit of small $R$, Eq.~(\ref{203}), the classical damping coefficient $\gamma$ in Eq.~(\ref{30}) should be substituted by $1/\tau_{0}$ in order to roughly estimate a part of the Lamb shift associated with an influence of the dielectric. It reads
\begin{equation}
\label{31}
\delta E^{(D)}_{L}\sim \frac{\hbar}{\tau_{0}}\sim V\left(\frac{\hbar\omega_{ph}}{V}\right)^{2}.
\end{equation}
One can conclude that when the atom is close to the dielectric, Eq.~(\ref{211}), the new contribution (\ref{31}) is of the order of the usual one
\begin{equation}
\label{32}
\delta E^{(D)}_{L}\sim\delta E_{L}\sim 10^{-6}V\sim 10^{10}\,{\rm s}^{-1}.
\end{equation}
So the anomalous Lamb shift is $\delta E_{L}+\delta E^{(D)}_{L}$ and one can expect in spectroscopic measurements an additional elevation of atomic levels when the atom is closer than $1000\,\AA$ to the dielectric surface.
\section{DISCUSSIONS}
\label{disc}
A new phenomenon is proposed in the paper. When an atom is separated by a macroscopic distance $R$ from a surface of a dielectric a formation of a quantum (bow) state in the dielectric occurs. The bow state of the system "atom above dielectric" is associated with a nonzero mean electric field ${\cal E}_{bow}$ produced by a polarization of the dielectric. Formation of that field is not due to classical effects as, for example, in the polaronic state when a lattice is classically polarized by an electron. Therefore the bow and the polaron are different objects.

When a dielectric has a flat superface, the coherence length $\xi$ of the system "atom above dielectric" increases with the distance $R$ between the atom and the dielectric. This occurs up to such $R$ when $\xi$ becomes of the size of the superface as in Fig.~\ref{fig6}. At a larger $R$ the bow state is not formed and the atom at that region has a lower energy compared to its position closer to the dielectric. For this reason the atom, moving from the large $R$, is reflected from the region of the smaller $R$ as in Fig.~\ref{fig6}.

When atoms cannot leave a region close to the dielectric, due  restricted geometry, for example, they will be organized into clusters to prevent bow formation and not to increase their energy. One can say that atoms become free being organized into clusters. This situation reminds quark confinement in particle physics since they are free at short distances (asymptotic freedom) \cite{POLYAKOV,LEE}.

Avoiding to occupy regions, with conditions for the bow state formation, is equivalent to some effective repulsion from those regions. This interaction mechanism is due to creation of a nonzero mean electric field. It substantially differs from the fluctuation mechanism when an energy of electromagnetic fluctuations (zero mean field) depends on a distance between bodies resulting in a force called the van der Waals force. In the bow state the mean nonzero value ${\cal E}_{bow}$ is extended over the coherence length and can be treated as the certain order parameter. According to Eq.~(\ref{11aaa}), the order parameter smoothly turns to zero under violation of bow sate conditions.

As pointed out in Sec.~\ref{interp}, the origin of the bow state is connected with violation of the interference of virtual overbarrier states of the atom (dephasing). One should make a more general remark about this phenomenon. For example, a transition from an incoherent radiation to a coherent one in lasers is a phase transition with formation of an order parameter in the coherent regime. In our case the situation is opposite. One should destroy interference of different virtual states by the dephasing mechanism to prevent their mutual compensation. The dephasing role is played by the relaxation processes. Since the electron moves under the barrier and its wave function drops down exponentially with distance we deal with processes of the underbarrier interference.

The problem, considered in the paper, is not a unique example of underbarrier interference. In two-dimensional motion under a barrier a smooth distribution of de Broglie waves, which mutually cancel each other, can be violated by formation of caustics which result in a strong dephasing effect \cite{IVLEV3}. In tunneling in two-dimensional Josephson system a strong multi-path interference is essential \cite{IVLEV4}.

Violation of the interference (dephasing) in disordered solids \cite{KHM,GANT} and in the bow state equally results in reduction of the localization. The difference is that in solids real electron states interfere but in the bow phenomenon virtual states participate in the interference. In the bow case reduction of the localization means a less decaying wave function.

In this paper we consider a single atom above the dielectric. For a gas of atoms the bow state does not disappear but becomes collective. This situation is to be studied in the future.

There is another aspect of the bow state. According to Eq.~(\ref{11aa}), a weak external electric field can destroy the bow state. The high frequency field, $\vec{\cal E}\cos\Omega t$, parallel to the dielectric surface destroys the bow state much more effectively since the cosine goes over into $\cosh\Omega\tau$  under the barrier \cite{MELN1,IVLEV1,IVLEV2}. This results in an exponentially enhanced effective electric field when the frequency $\Omega$ is larger than an inverse characteristic time of the problem. For example, the frequency of He-Ne laser, $\Omega\sim 10^{15}\,{\rm s}^{-1}$, satisfies that condition. So a weak laser beam, in the direction perpendicular to the dielectric surface, can destroy the bow state. A switch on time of the laser can be easily taken as $\Delta t\sim 10^{-10}$\,s.

We proposed an anomalous Lamb shift, $\delta E_{L}+\delta E^{(D)}_{L}$, of atomic energy levels when the atom is close to the dielectric. The contribution $\delta E_{L}$ is due to the usual mechanism when a single atom interacts with electromagnetic fluctuations in the infinite space. The second one, $\delta E^{(D)}_{L}$, is related to electromagnetic fluctuations in the system "atom above dielectric". The both contributions are of the same order of magnitude.

Underbarrier interference and dephasing are always counterintuitive phenomena resulting in unexpected consequences. Easy penetration through classical potential barriers is possible due to the dephasing of overbarrier waves \cite{IVLEV3}. Tunneling in Josephson junctions via interfering multiple paths results in different physical properties of junctions \cite{IVLEV4}. A study of underbarrier interference in quantum mechanics and other wave processes is worth to be continued.
\section{CONCLUSION}
A new manifestation of underbarrier dephasing is proposed in addition to the previous cases when it was realized in two-dimensional tunneling. The atomic electron, due to the interaction with other atoms, undergoes virtual transitions to the continuous spectrum. The compensating interference of the propagating waves does not allow an electron wave function other than an exponentially decayed one. Under effects of dissipation on the electron motion the compensating interference of the propagating waves gets reduced due to dephasing and the electron state becomes at large distances a superposition with a fraction of overbarrier waves. This is equivalent to acceleration of the electron by some mean electric field which is formed in the system. The state with the mean electric field is called the bow state.

\end{document}